\documentclass[11pt,twoside]{article}

\usepackage{asp2006}
\usepackage{epsf}
\usepackage{psfig}
\usepackage{lscape}

\begin{document}
\title{The {\it GALEX} UV emission in shell galaxies}   
\author{A. Marino\altaffilmark{1}, R. Rampazzo\altaffilmark{1}, R. Tantalo\altaffilmark{2}, D. Bettoni\altaffilmark{1}, L. M. Buson\altaffilmark{1},  C. Chiosi\altaffilmark{2}, G. Galletta\altaffilmark{2}}   
\affil{{1} INAF - Osservatorio Astronomico di Padova, Vicolo dell'Osservatorio 5, 
I-35122 Padova, Italy}   
\affil{{2} Dipartimento di Astronomia Universit\`a di Padova, 
 Vicolo dell'Osservatorio 2, 35122 Padova (Italy)}

\begin{abstract} 
Shell galaxies are widely considered the debris of recent accretion/merging
episodes. Their high frequency in low density environment suggests that such
episodes could be among the driver of the early-type galaxy secular evolution. 
We present far and near UV  (FUV and NUV  respectively hereafter)  {\it GALEX}
 photometric properties  of a sample of shell galaxies.  
\end{abstract}

\section{Introduction} 
In a hierarchical evolutionary scenario, galaxies experience accretion/merging
events during their lifetime. While early-type  galaxies in nearby clusters appear
(homogeneously) old, the field early-type galaxy population seems to contain 
genuinely, recently {\it rejuvenated} objects \citep[see e.g.][]{Clemens06}.
Early-type galaxies showing fine structure, like shells, occupy a special position
since they are believed to fill the gap between ongoing mergers and normal 
elliptical galaxies.
The UV emission is crucial to test whether these galaxies do host ongoing/recent star
formation processes and study  their distribution across the galaxy. 
We present  new  {\it GALEX} observations of  three  shell galaxies NGC 1210, 
MGC -05-07-1 (GI04-0030-0059 PI D. Bettoni) and NGC 5329 (from archive) in addition 
to those analyzed in \citet{Rampazzo07} which we use as baseline for our preliminary conclusions.
\section{UV data and discussion}
Table 1 summarizes the journal and the basic results of the {\it GALEX} observations.

\begin{table*}
\scriptsize
\caption{Journal of the {\it GALEX} observations}
\begin{tabular}{lcclllll}
 \hline
 \multicolumn{1}{l}{Name}&
\multicolumn{1}{c}{NUV}  &
\multicolumn{1}{c}{FUV}  &
 \multicolumn{1}{l}{P.I.}&
\multicolumn{1}{c}{m$_{NUV}^{tot}$}&
\multicolumn{1}{c}{m$_{FUV}^{tot}$} &
\multicolumn{1}{c}{FUV-NUV}\\
\multicolumn{1}{c}{}&
\multicolumn{1}{c}{exposure [sec]} &
\multicolumn{1}{c}{exposure [sec]} &
 \multicolumn{1}{c}{}&
\multicolumn{1}{c}{}&
\multicolumn{1}{l}{}&
\multicolumn{1}{l}{}\\
\hline
 MCG-05-07-1  &1510 & 1531   & D. Bettoni & 18.67$\pm$0.07 & 19.76$\pm$0.07 & 1.11$\pm$0.10  \\
NGC 1210  &1558 & 1608 &  D. Bettoni &17.14$\pm$0.02&20.08$\pm$0.07&2.95$\pm$0.07  \\
NGC5329  &3889  &2666   &   MIS&18.20$\pm$0.02&20.20$\pm$0.04&2.02$\pm$0.05 \\
\hline
\end{tabular}
\label{table1}
\medskip{FUV - NUV AB magnitudes have been corrected for Galactic extinction.}
\end{table*}
Smoothed images and 2D colour maps are shown in Figure 1. FUV emission in NGC 5329  
 is present only in the central part of the galaxy. 
In NGC 1210 and MGC-05-07-01 the FUV is quite strong both in the polar ring of the 
former galaxy and in the debris systems, residual of the accretion events of  both galaxies. 
The (FUV-NUV) colour in the  north tail of  NGC 1210  ($\sim$0.31, $\sim$0.34, $\sim$0.96), in the  polar ring   ($\sim$0.36, $\sim$0.73) and in the nucleus ($\sim$1.11) of MGC-05-07-01 are quite blue.  \citet{Neff05} show that tail/bridges produced by interaction may have similar colour indicative of a recent star formation (200-300 Myr).  
 The age of 1-3 Gyr estimated by Whitmore et al. (1987),   responsible for the present structure of MCG-05-07-01, is consistent with age indication coming from (UV - optical) colours. 
\begin{figure}
\centerline{\psfig{figure=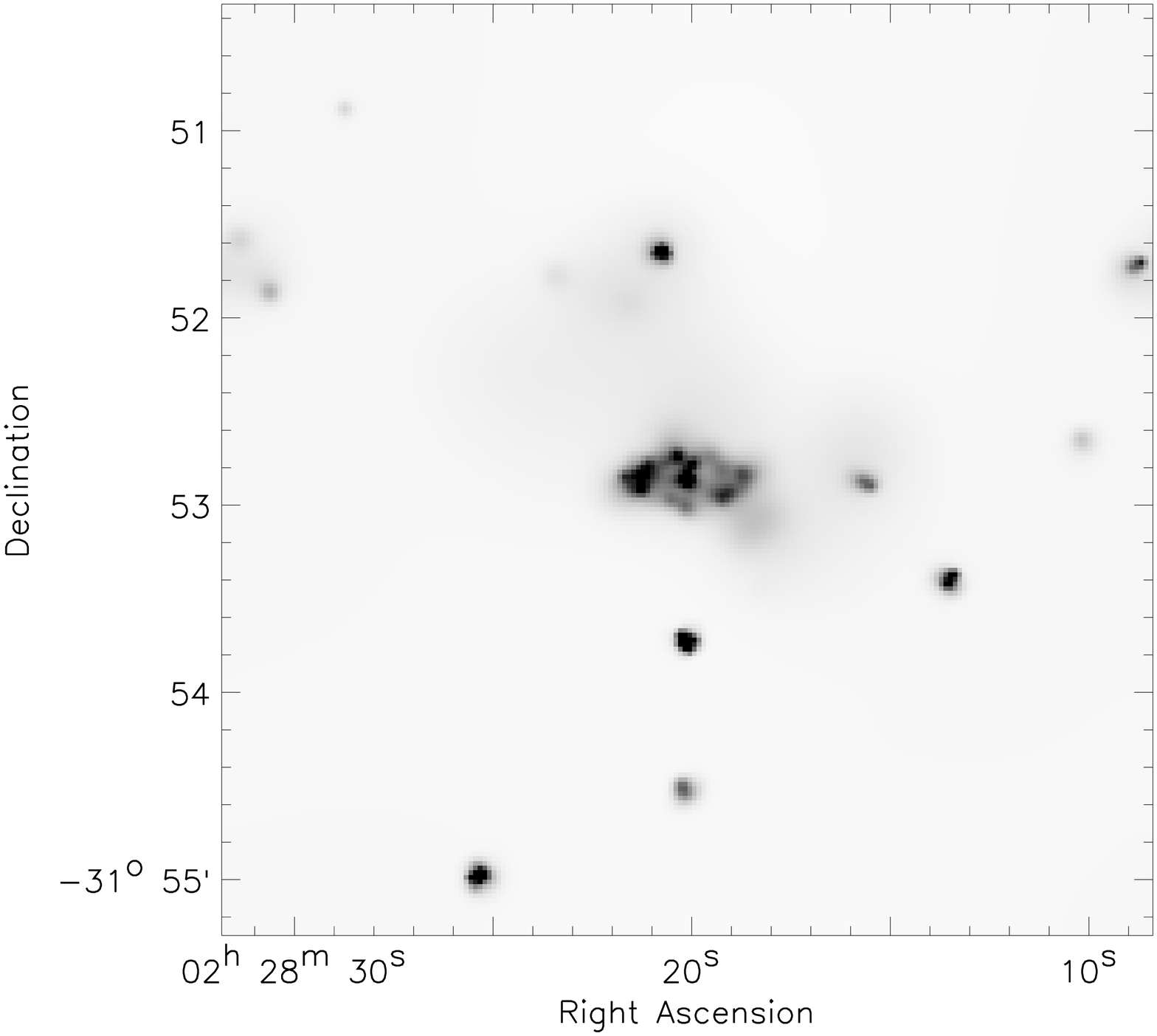,width=2.8cm}\psfig{figure=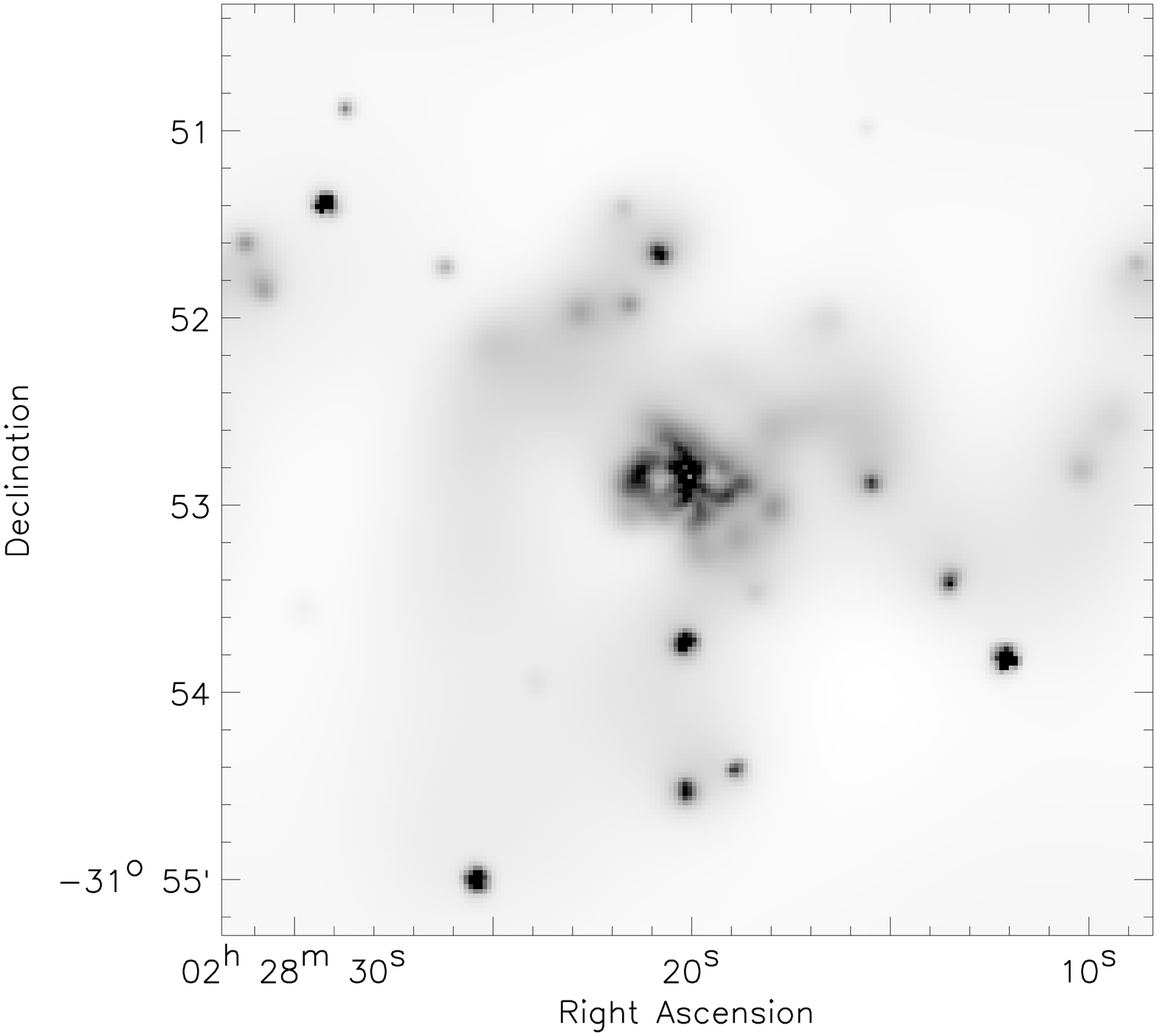,width=2.8cm}}
\centerline{\psfig{figure=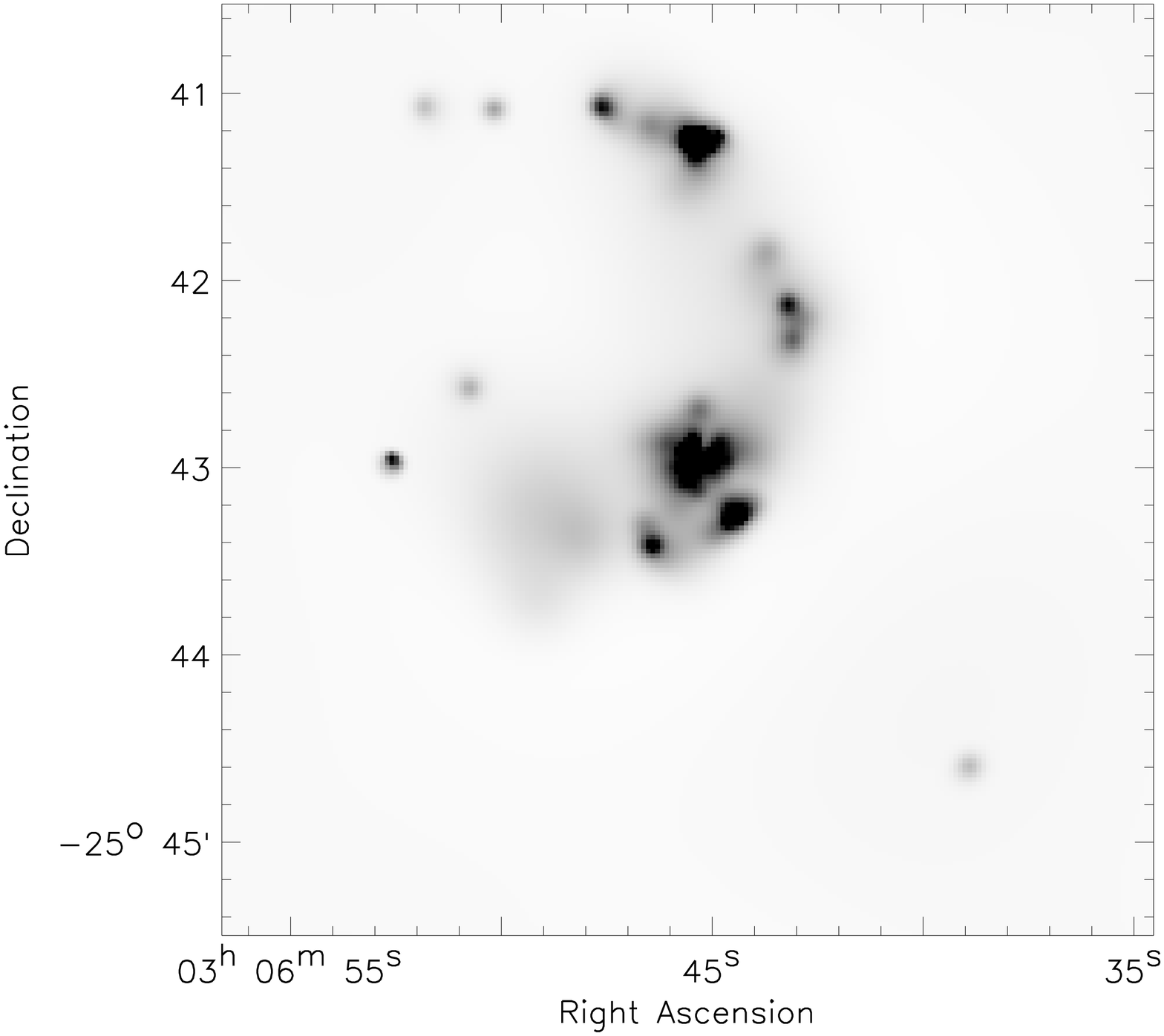,width=2.8cm}\psfig{figure=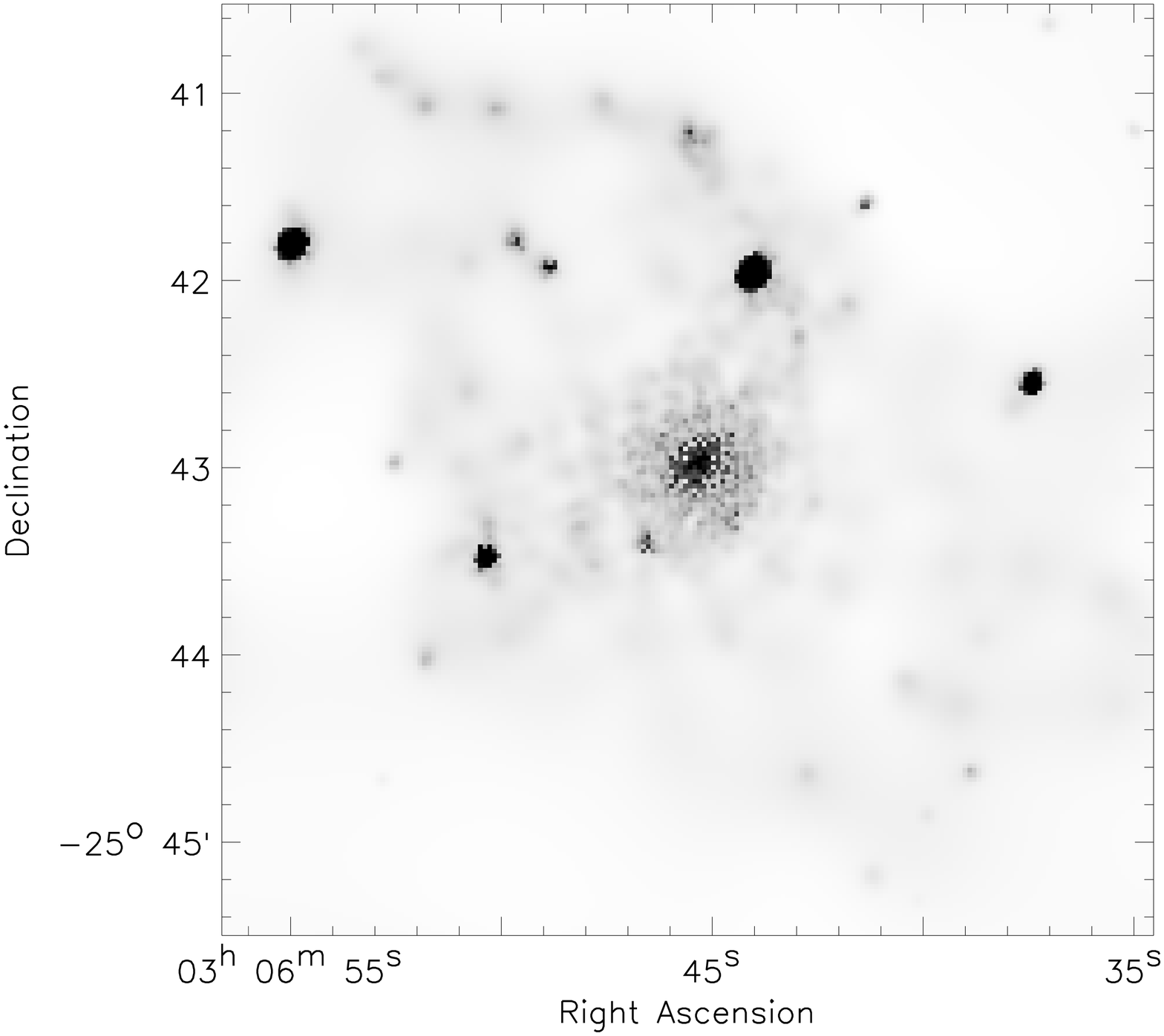,width=2.8cm}}
\centerline{\psfig{figure=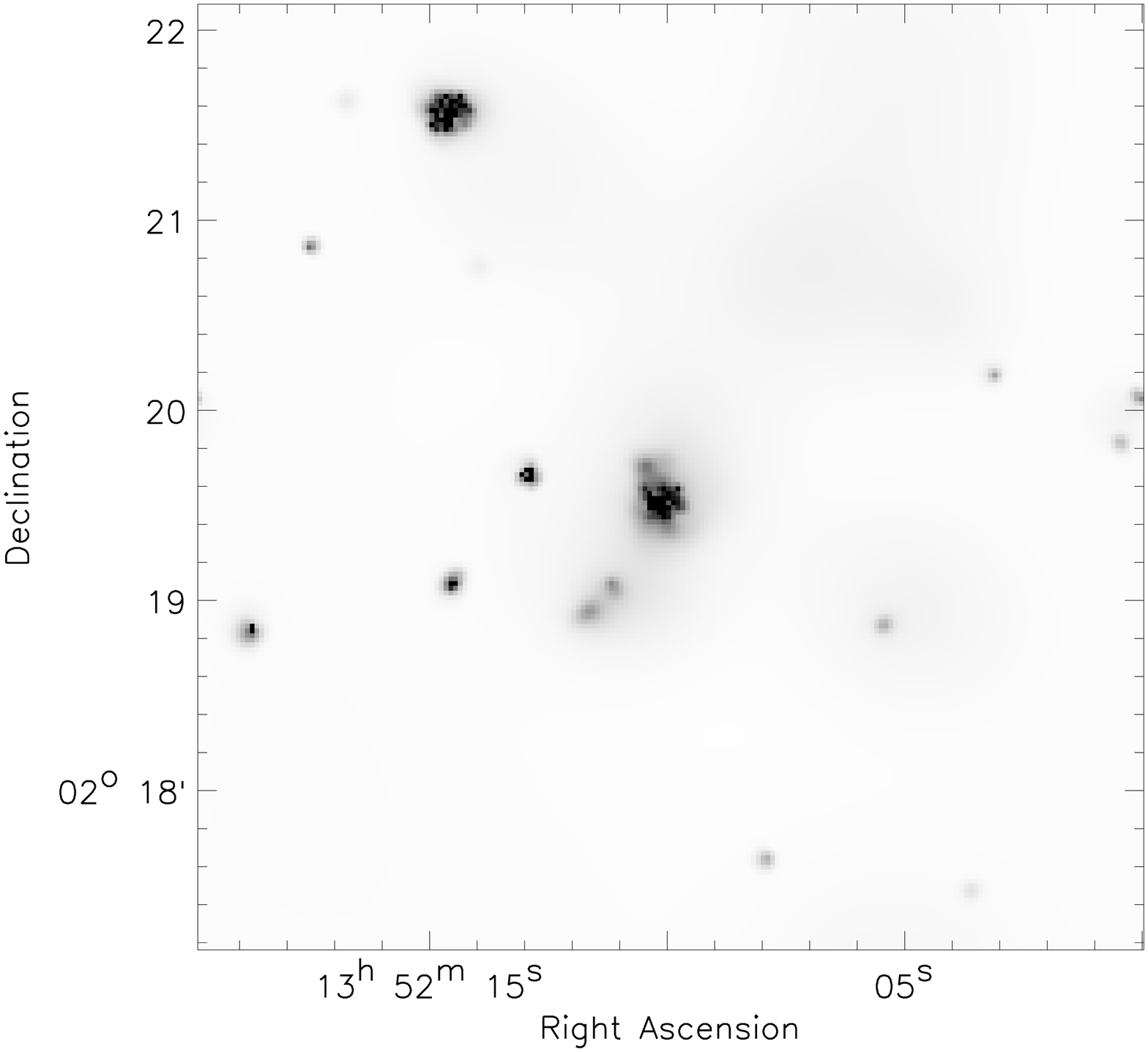,width=2.8cm}\psfig{figure=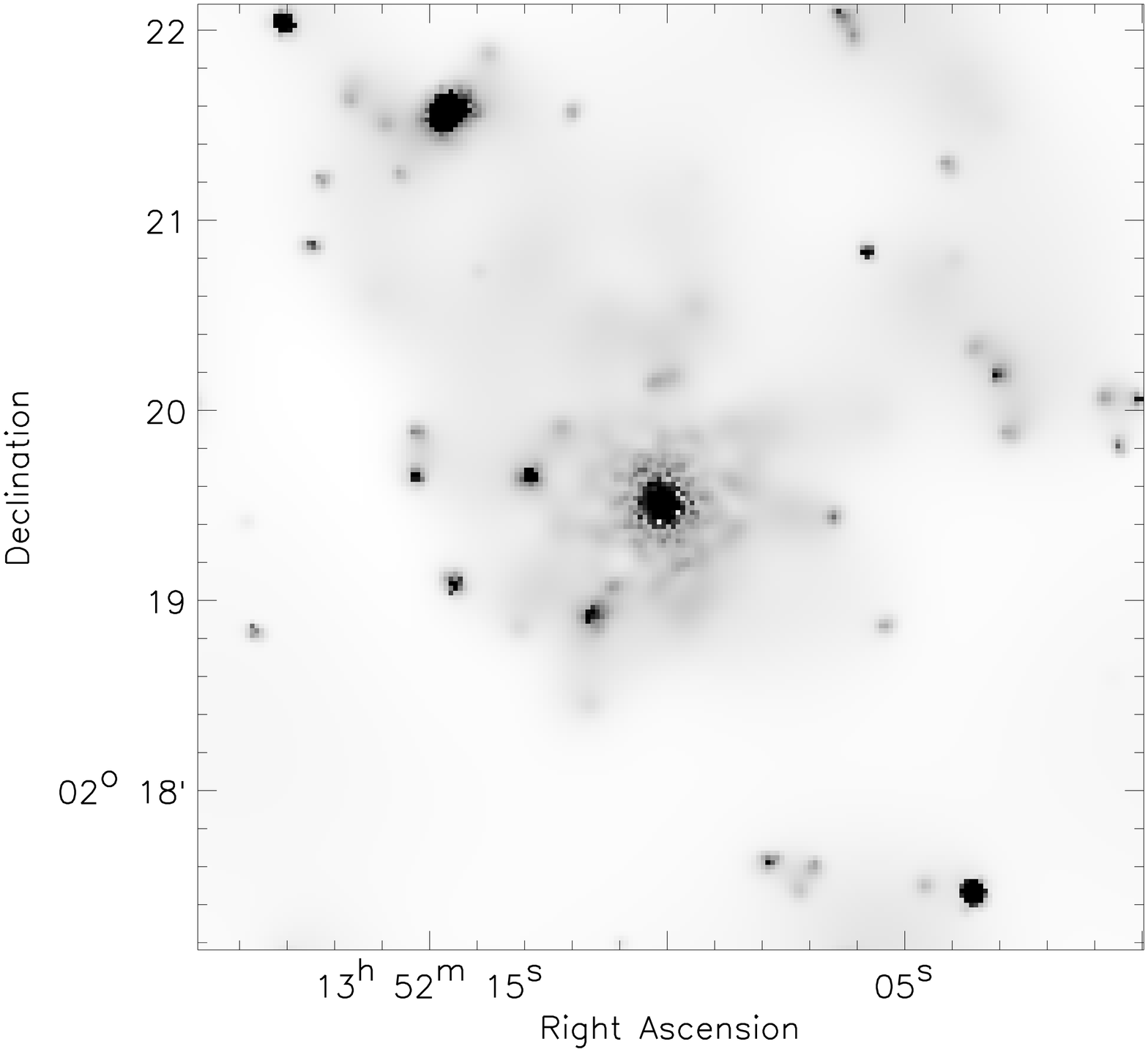,width=2.8cm}}
\caption{{\it GALEX} FUV (left panels), NUV (middle panels)  images  (5\arcmin $\times$ 5\arcmin) and (FUV-NUV) colour maps (right panels) of MCG-05-07-1 (top row), NGC 1210 (mid row) and NGC 5329 (bottom row).  FUV  and NUV  images are smoothed using {\it Asmooth} \citep{Ebeling06}.}
\label{fig1}
\end{figure}
\acknowledgements 
This research has been partially founded by ASI-INAF contract I/023/05/0.
{\it GALEX} is a NASA Small Explorer, operated for NASA by California Institute of 
technology under NASA contract NAS-98034.

\end{document}